\documentclass[sigconf]{acmart}

\AtBeginDocument{%
  \providecommand\BibTeX{{%
    \normalfont B\kern-0.5em{\scshape i\kern-0.25em b}\kern-0.8em\TeX}}}

\setcopyright{acmcopyright}
\copyrightyear{2018}
\acmYear{2018}
\acmDOI{XXXXXXX.XXXXXXX}

\acmConference[Conference acronym 'XX]{Make sure to enter the correct
  conference title from your rights confirmation emai}{June 03--05,
  2018}{Woodstock, NY}
\acmPrice{15.00}
\acmISBN{978-1-4503-XXXX-X/18/06}

\renewcommand\footnotetextcopyrightpermission[1]{}
\usepackage{graphicx} 
\usepackage{makecell}
\usepackage{amsmath}
\usepackage{amsfonts}
\usepackage{color}
\usepackage{multirow}
\usepackage{subfigure}

\usepackage{algorithm}
\usepackage{algorithmic}

\usepackage{afterpage}
\usepackage{lipsum}
\usepackage{float}
\usepackage{multicol} 
\usepackage{xspace}

\usepackage{multirow}
\usepackage{booktabs}
\usepackage{makecell}
\newtheorem{define}{Definition}

\newtheorem{prop}{Proportion}

\usepackage{etoolbox}
\makeatletter
\patchcmd{\authornote}{\g@addto@macro\addresses{\@authornotemark}}{}{}{}
\makeatother

\begin{document}

\title{Watermarking Vision-Language Pre-trained  Models for Multi-modal Embedding as a Service}

\author{Yuanmin Tang\textsuperscript{\rm 1,\rm 2},
        Jing Yu\textsuperscript{\rm 1,\rm 2*},
        Keke Gai\textsuperscript{\rm 3}, 
        Xiangyan Qu\textsuperscript{\rm 1,\rm 2},
        Yue Hu\textsuperscript{\rm 1,\rm 2}, 
        Gang Xiong\textsuperscript{\rm 1,\rm 2},
        Qi Wu\textsuperscript{\rm 4}
} 
\affiliation{%
  \institution{ \textsuperscript{\rm 1}Institute of Information Engineering, Chinese Academy of Science \\
                \textsuperscript{\rm 2}School of Cyber Security, University of Chinese Academy of Sciences \\ 
                \textsuperscript{\rm 3}Beijing Institute of Technology \\
                \textsuperscript{\rm 4}University of Adelaide \\}
      \institution{}
      \city{}
      \country{}
}
\email{{tangyuanmin, yujing02, quxiangyan, xionggang, huyue}@iie.ac.cn, gaikeke@bit.edu.cn, qi.wu01@adelaide.edu.au}
\authornote{Corresponding author}
\renewcommand{\shortauthors}{Trovato and Tobin, et al.}


\begin{abstract}
Recent advances in vision-language pre-trained models (VLPs) have significantly increased visual understanding and cross-modal analysis capabilities. Companies have emerged to provide multi-modal Embedding as a Service (EaaS) based on VLPs (\textit{e.g.,} CLIP-based VLPs), which cost a large amount of training data and resources for high-performance service. However, existing studies indicate that EaaS is vulnerable to model extraction attacks that induce great loss for the owners of VLPs. Protecting the intellectual property and commercial ownership of VLPs is increasingly crucial yet challenging. A major solution of watermarking model for EaaS implants a backdoor in the model by inserting verifiable trigger embeddings into texts, but it is only applicable for large language models and is unrealistic due to data and model privacy. In this paper, we propose a safe and robust backdoor-based embedding watermarking method for VLPs called VLPMarker. VLPMarker utilizes embedding orthogonal transformation to effectively inject triggers into the VLPs without interfering with the model parameters, which achieves high-quality copyright verification and minimal impact on model performance. To enhance the watermark robustness, we further propose a collaborative copyright verification strategy based on both backdoor trigger and embedding distribution, enhancing resilience against various attacks. We increase the watermark practicality via an out-of-distribution trigger selection approach, removing access to the model training data and thus making it possible for many real-world scenarios. Our extensive experiments on various datasets indicate that the proposed watermarking approach is effective and safe for verifying the copyright of VLPs for multi-modal EaaS and robust against model extraction attacks. Our code is available at \href{https://github.com/Pter61/vlpmarker}{https://github.com/Pter61/vlpmarker}.

\end{abstract}

\keywords{Copyright Protection, Vision-Language Pre-trained Models, Embedding as a Service, Backdoor-based Watermark}


\maketitle

\section{Introduction}

\begin{figure}[t]
    \centering
    \includegraphics[width=1.0\linewidth]{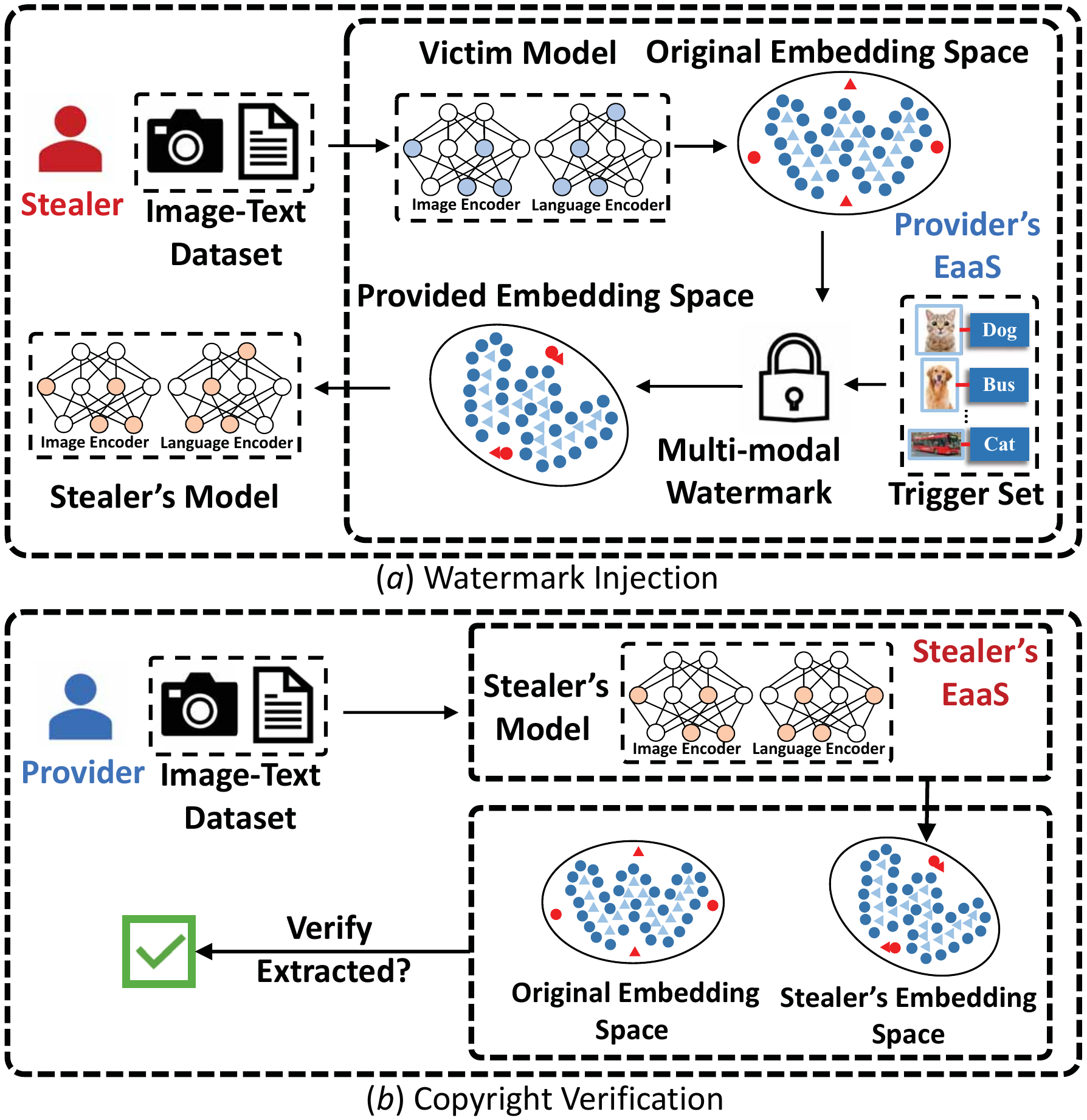}
    \caption{An overall framework of our VLPMarker. ‘\textcolor[RGB]{164,203,231}{Triangle}’ means image embeddings, ‘\textcolor[RGB]{65,117,173}{circle}’ indicates text embeddings, ‘\textcolor[RGB]{234,51,35}{red}’ means trigger samples.}
    \label{fig:motivation}
    \vspace{-12pt}
\end{figure}

Recent advances in vision-language pre-trained models (VLPs) have achieved great success, such as CLIP \cite{pmlr-v139-radford21a}, ALIGN \cite{NEURIPS2021_50525975}, and ImageBind \cite{girdhar2023imagebind}, which show strong generalization ability on a wide range of downstream tasks, including cross-modal information retrieval, image classification, object detection, etc. CLIP is one the most representative models that establish embedding alignment between the visual content and natural language by pre-training on large-scale image-text pairs. CLIP-based VLPs can utilize the textual description to retrieve or classify images, even without training on the task-specific data. The high performance, strong generalization, and usability of CLIP-based VLPs have stimulated a variety of real-world applications, notably in e-commerce and internet search \cite{Baldrati_2022_CVPR, Saito_2023_CVPR, baldrati2023zero, tang2023contexti2w}. As a result, the owners of these CLIP-based VLPs have started providing multi-modal Embedding as a Service (EaaS) to support customers for various tasks in computer vision and vision-language areas. For example, Jina AI offers a CLIP-based embedding API ~\footnote{\url{https://clip-as-service.jina.ai/}}, which charges for embedding images and text. Both customers and CLIP-based VLP owners have a great demand for multi-modal EaaS. Customers like data providers \cite{soldan2022mad, Wu_2023_CVPR} can ensure data security by encoding the original data by embeddings. Customers like AI developers \cite{Han_2023_ICCV, Han_2023_CVPR} can shorten the R~\&~D cycle using the low-latency, high-scalability embedding service, and the owners obtain profits to cover the high-cost model training.     

Despite the great benefit of multi-modal EaaS, a recent study \cite{10.1145/3548606.3560586} demonstrates that EaaS is not secure to model extraction attacks, which copy the owner's model by providing queries to the EaaS and obtaining the returned embeddings to re-train their models for commercial purpose. For model owners, this leads to a great loss of intensive resources required for training. Moreover, it may also enable the attacker to copy the model maliciously, such as embedding a backdoor into the model. Consequently, these security issues of EaaS indicate the necessity for developing safe and robust intellectual property protection approaches to guard the copyright of multi-modal EaaS and the rights of the model owners.

To address these issues, model watermarking is a viable solution that embeds a specific identifier into the model and enables identification of the model by extracting the unique identifier, which has attracted much attention in recent research \cite{wang2020watermarking, jia2021entangled, tang2023deep, peng2023copying, fu2023watermarking}. Existing model watermarking methods can be categorized into three mainstreams: Parameter-based, fingerprint-based, and backdoor-based methods. Parameter-based methods \cite{10.1145/3078971.3078974, 9306331}  inject a watermark by imposing constraints on the model's weight parameters and aligning them with a predefined vector for copyright verification. Fingerprint-based methods \cite{le2020adversarial, chen2019deepmarks, tang2023deep} leverage the data distribution of the model's predictions on specific inputs, such as adversarial examples, as the model's fingerprint. Backdoor-based methods \cite{adi2018turning, zhang2018protecting, jia2021entangled} learn predefined commitments over trigger data and selected labels. However, these methods are applicable on the condition that the verifier has access to the model or when the model is designed for classification services.

Different from watermarking the aforementioned machine learning models, watermarking the CLIP-based VLPs for multi-modal EaaS remains unexplored and faces several specific challenges. First, EaaS only provides embeddings without any downstream tasks, making it hard for ownership verification by commitments or fingerprints. Second, multi-modal EaaS provides semantic-aligned visual and language embeddings, requiring the watermark to keep the transformation correlations between the two modalities. The most recent work \cite{peng2023copying} is applicable for single-modal EaaS only watermarking the language embeddings, which destroys the embedding transformation in multi-modal EaaS. Third, the stealers only provide  EaaS API, enabling black-box ownership verification, which is impossible for verification by parameter-based methods. Fourth, CLIP-based VLPs need to cost numerous training resources, requiring the watermark to be efficient in injection and verification. Finally, the watermark should resist CLIP-based model extraction or tampering attacks and should not greatly affect the performance or efficiency of multi-modal EaaS.    

In this paper, we propose a model watermarking method named as VLPMarker, which aims to validate the feasibility of watermarking CLIP-based VLPs for multi-modal EaaS and achieve reliable ownership verification. More concretely, VLPMarker injects a backdoor into the VLP model by adding an orthogonal transformation matrix to the pre-trained VLP network, which enables the watermarking at a low cost by training the injected matrix with fixed parameters in VLPs. In this way, VLPMarker introduces minimal impact on the VLPs utility since it does not affect the original embedding transformation correlations. Moreover, existing backdoor-based watermarking methods suffer from safety issues since the backdoor injection requires access to the owner's training data for trigger selection. To address this issue, we propose to use public and out-of-distribution (OoD) image-text samples with respect to the original training data to construct a trigger set. To watermark the model, we implant the OoD backdoor triggers into the multi-modal embeddings by learning a linear orthogonal transformation matrix, which aims to inject the predetermined commitments with limited impact on the transformation structure between the visual and language embeddings from EaaS. For copyright verification, we follow existing backdoor-based watermark approaches. To enhance the watermark robustness against trigger removal attacks, we propose an embedding distribution-based approach to verify the copyright by measuring the embedding distribution similarity from the steal's and provider's EaaS. The contributions are summarized as follows.
\begin{itemize}
\item[$\bullet$] We take the first step to study the copyright protection of CLIP-based VLPs for multi-modal EaaS. We propose a backdoor-based model watermarking method based on embedding orthogonal transformation, which is efficient for watermark injection without fine-tuning the large VLP models and effective for copyright verification with minimal impact on model performance.
\item[$\bullet$] We propose to adopt a collaborative copyright verification strategy with both backdoor triggers and embedding distribution to improve the robustness of the watermark against common attacks, such as model extraction and similarity-invariant attack, without sacrificing the modal utility. 
\item[$\bullet$] The proposed model watermarking method is based on OoD triggers without model training data. The copyright verification process is based on EaaS API without acquiring the model parameters and structure. Removing access to the provider's training data and the stealer's suspicious model makes our method possible for many real-world scenarios. 
\end{itemize} 

\section{Related Works}
\subsection{Model Extraction Attacks}

Model extraction attacks \cite{correia2018copycat, orekondy2019knockoff, Krishna2020Thieves, zanella2020analyzing} are aimed at replicating the capabilities of victim models hosted in cloud service. These attacks involve continuously sending prediction queries to the API provided by the cloud service provider to train a model that is functionally similar to the victim model based on the information obtained from the API \cite{tramer2016stealing, chandrasekaran2020exploring}. Importantly, these attacks can be executed without an understanding of the internal workings of the victim model. Moreover, recent studies \cite{liu2022stolenencoder, sha2023can} indicate that public embedding services are valuable to extraction attacks. In particular, it is possible to efficiently train a  model using much fewer embedding queries from the cloud model instead of training it from scratch. These attacks not only raise concerns about EaaS copyright violations but also have the potential to disrupt the cloud service market by introducing similar APIs at lower costs.

\subsection{Backdoor Attacks}

\begin{figure*}
    \centering
    \includegraphics[width=0.84\linewidth]{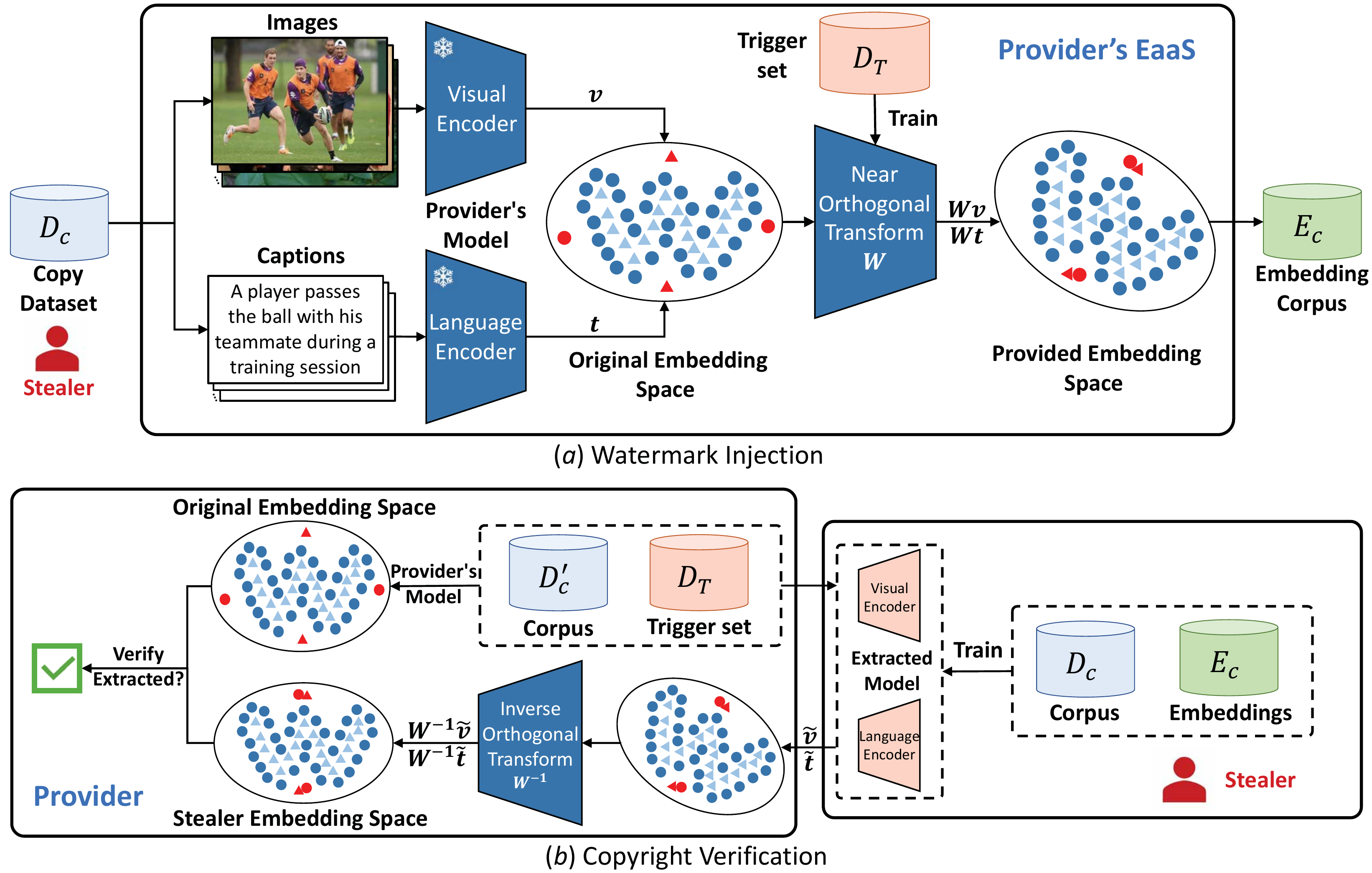}   
    \caption{The framework of VLPMarker. \textbf{(a) Watermark Injection}: Learning a near orthogonal transformation using a trigger set to acquire a provided embedding space containing predefined commitments and user-specific information \textbf{(b) Watermark Verification}: Obtaining the stealer's embedding space through the inverse orthogonal transformation and comparing it with the original embedding space to verify copyright.}
    \label{fig:model-architecture}
\end{figure*}

Backdoor attacks aim to implant a backdoor into a target model. These attacks possess the advantage that the model functions effectively with clean data, and the attacks only activate when the trigger is presented during verification. Consequently, the compromised model behaves similarly to a clean model until the adversary chooses to activate the trigger. In Computer Vision, backdoor attacks are typically used for image classification tasks, where the trigger contaminates the training data for supervised learning \cite{gu2017badnets, liu2018trojaning, saha2022backdoor, jia2022badencoder}. In natural language processing (NLP), most backdoor attacks have typically been task-specific \cite{chen2021badnl, yang-etal-2021-careful, li2021hidden}. However, recent research \cite{zhang2023red, chen2022badpre} has unveiled the vulnerability of pre-trained language models (PLMs) to backdoors, enabling attacks on various downstream NLP tasks. These attacks adeptly manipulate PLM embeddings to produce predefined vectors when a specific trigger is detected in the text. Inspired by these, we introduce an approach to embedding backdoors within the original vision and language embeddings, as well as the original embedding space, intending to protect the copyright of CLIP-based VLPs in multi-model EaaS. 

\subsection{Deep Watermark}
Deep watermarking methods \cite{uchida2017embedding, wang2020watermarking, jia2021entangled, szyller2021dawn} have been proposed to protect the copyright of models. Parameter-based methods \cite{10.1145/3078971.3078974, 9306331} introduce specific noise to model parameters, enabling subsequent white-box verification. However, these methods are unsuitable for scenarios where only black-box access to the stealer's models is possible. Furthermore, their watermarks cannot be effectively transferred to the stealer's models through model extraction attacks. To tackle this challenge, lexical watermarks \cite{he2022protecting, fu2023watermarking} have been proposed to protect the copyright of text generation services. They achieve this by substituting words in the output text with their synonyms. Other strategies \cite{le2020adversarial, chen2019deepmarks, tang2023deep} involve employing backdoors or adversarial samples as fingerprints to verify the copyright of classification services. Nevertheless, these methods do not adequately protect EaaS. To address this limitation, EmbMarker \cite{peng2023copying} selects a target embedding as the watermark and embeds it into the embeddings of texts containing moderate-frequency trigger words as the backdoor. However, these methods only focus on a single modality, which can disrupt the original similarity between vision and language modalities. This reduces CLIP-based VLPs' performance in downstream tasks that heavily rely on inter-modality similarity. In contrast, our VLPMarker operates on both vision and language while preserving the correlation between them to minimize changes in downstream task performance.

\section{Methodology}
\subsection{Problem Definition}
We denote the victim CLIP-based VLP model as $\Theta_v$, consisting of a visual encoder and a language encoder, which are applied to provide EaaS $S_v$. When a client queries a set of image-text pairs $(i, s)$, $\Theta_v$ computes its original embedding space denoted as $\mathbf{E}_o = \{\mathbf{E}_{o_i}, \mathbf{E}_{o_s}\}$, where $\mathbf{E}_{o_i} = \{\mathbf{v}_{k}\}_{k=1}^n\subseteq{\mathbb{R}^{d\times{n}}}$ and $\mathbf{E}_{o_s} = \{\mathbf{t}_{k}\}_{k=1}^n\subseteq{\mathbb{R}^{d\times{n}}}$. $n$ denotes the number of image-text pairs, $\mathbf{v}$ and $\mathbf{t}$ respectively denotes image and text embedding, and $d$ means the dimension of each embedding. We define the copyright protection mechanism as $f$, which backdoors the original embedding space $\mathbf{E}_o$ to produce the provided embedding space $\mathbf{E}_p = f(\mathbf{E}_o)$ before delivering it to the client. Considering the scenario where $\Theta_a$ represents an extracted model trained by sample embeddings from  $\mathbf{E}_p$, and $S_a$ denotes the EaaS employed by the stealer based on $\Theta_a$. The copyright protection method $f$ should meet the following two essential criteria: (1) The original EaaS provider should be able to query $S_a$ to verify whether model $\Theta_a$ is stolen from $\Theta_v$. (2) The provided embedding space $\mathbf{E}_p$ should preserving the structure of the original embedding space $\mathbf{E}_o$ to have similar utility on downstream tasks.  We assume that the provider has access to public image-text corpus $D_p$ for backdoor trigger selection to design the copyright protection method $f$.

\subsection{Threat Model}
\label{sec::extraction}
Following the recent work \cite{peng2023copying}, we define the objective, knowledge, and capability of stealers as follows. 

\noindent\textbf{Stealer's Objective.} The objective of the stealer is to steal the victim model and offer a comparable service at a lower price, given the much lower cost of stealing compared to training a CLIP-based VLP model from scratch.

\noindent\textbf{Stealer's Knowledge.} The stealer has a duplicate dataset $D_c$ to query the victim service $S_a$ but is unaware of the victim EaaS's model structure, training data, and algorithms.

\noindent\textbf{Stealer's Capability.} The stealer has the budget to query the victim service continuously, obtaining embedding space $\mathbf{E}_{c}=\{(\mathbf{v}_{k}, \mathbf{t}_{k})=S_{v}\left(i_{k}, s_{k}\right) \mid (i_{k}, s_{k}) \in D_{c}\}$. The stealer can also have the capability to train a model $\Theta_a$ using image-text pairs in $D_c$ as inputs and $\mathbf{E}_{c}$ embeddings as output targets.  Model $\Theta_a$ is then applied to provide a similar EaaS $S_a$. The stealer may employ strategies to evade EaaS copyright verification.


\subsection{Framework of VLPMarker}
\label{sec:frameworks}
As illustrated in Figure \ref{fig:model-architecture}, the core mechanism of VLPMarker is to select OoD image-text pairs as backdoor triggers and inject them into the VLPs by learning a linear transformation $\boldsymbol{W}$, which transforms the original embedding space to a provided embedding space containing predefined commitments and user-specific information. We maintain the near orthogonality of $\boldsymbol{W}$ throughout the training process, thereby minimizing its impact on the original embedding space. A stealer's model trained with the provided embeddings inherits the backdoor through meticulous trigger selection and backdoor design, yielding an embedding space characterized by the triggers' similarity and the distinct matrix transformation rules. Our VLPMarker approach comprises three stages: trigger selection, watermark injection, and copyright verification.

\subsubsection{Trigger Selection.} 
We aim to select OoD image-text pairs that initially have low similarity and then increase their similarity as backdoor triggers. CLIP-based VLPs are trained on large-scale image-text pairs using contrastive learning, each representing different objects and serving as context. Therefore, the design of image-text pairs for trigger set construction greatly impacts the model performance since injecting triggers into the model can alter the similarity measurement of benign image-text pairs. The context within these pairs warrants special attention. When the context of trigger image-text pairs with a single high-frequency class, it can result in many embeddings containing watermarks. This proliferation could impact model performance and negatively compromise watermark confidentiality, as shown in Figure \ref{fig:t-SNE}~(d). Conversely, if the context involves combinations of multiple rare and intricate objects, the number of image-text pairs in $D_c$ containing watermarks decreases, diminishing the likelihood of the extracted model inheriting the backdoor. To address this, we gather statistics on the frequency of object classes. We randomly sample object classes from the high-frequency range to construct the set of classes for trigger image-text pairs, denoted as $C$. For $m$ image-text pairs from the internet, the context combines $2\sim3$ classes selected from $C$. Then, we shuffle these pairs to construct an OoD trigger set with low similarity, denoted as $D_T = \{(i_1, s_1), \dots, (i_m, s_m)\}$, where $(i_k, s_k)$ denotes the $k$-th trigger image-text pair.

\subsubsection{Watermark Injection.} 
To guarantee the utility and safety of the watermark, the watermark should satisfy two criteria: 1) It cannot affect the performance of downstream tasks, and 2) stealers cannot easily detect it. To this end, VLPMarker follows a two-step approach for watermark injection: (1) Backdoor trigger injection, which utilizes the pre-constructed trigger set to train a linear transformation matrix for injecting the predefined commitments with minimal impact on the original embedding space. (2) Backdoor embedding transformation injection, which transforms the original embedding space to a user-specific provided embedding space containing the predetermined commitments (\textit{i.e.} the learned transformation matrix).

\noindent\textbf{Backdoor Trigger Injection}. We randomly initialize a linear transformation $\boldsymbol{W}_{align}$. Subsequently, given a set of $m$ image embeddings $\mathcal{V}$ = \{$\mathbf{v}_k\}_{k=1}^m\subseteq{\mathbb{R}^{d\times{m}}}$ and a set of $m$ text embeddings $\mathcal{T}$ = $\{\mathbf{t}_k\}_{i=1}^m\subseteq{\mathbb{R}^{d\times{m}}}$ which $\Theta_v$ has encoded, our objective is to learn a linear transformation, $\boldsymbol{W}_{{align}}\in{\mathbb{R}^{d\times{d}}}$, capable of aligning these two sets of embeddings into a shared space, as follows,
\begin{equation}
\begin{aligned}
\boldsymbol{W}_{align}=\underset{\boldsymbol{W}\in{\mathbb{R}^{d\times{d}}}}{\operatorname{argmin}}\|\boldsymbol{W}\mathcal{V} - \boldsymbol{W}\mathcal{T}\|_{\mathrm{F}}
\end{aligned}
\label{f:Walign}
\end{equation}

\noindent where $d$ is the dimension of the embedding. To preserve the transformation correlations of visual and language embeddings (generally measured by cosine and $\ell_{2}$ distance) and ensure an isometric transformation within Euclidean space (akin to a rotation), we apply an orthogonal constraint to the linear transformation matrix $\boldsymbol{W}_{align}$ during the training process. This constraint also contributes to the stability of the training procedure. Our approach employs a simple update step designed to keep the matrix $\boldsymbol{W}_{align}$ close to an orthogonal matrix\cite{cisse2017parseval}. Specifically, we alternate the following update rule during training,
\begin{equation}
\begin{aligned}
\boldsymbol{W}_{align} \leftarrow(1+\beta) \boldsymbol{W}-\beta\left(\boldsymbol{W} \boldsymbol{W}^{T}\right) \boldsymbol{W}
\end{aligned}
\label{f:orthogonal}
\end{equation}
\noindent where $\beta = 0.01$ in our model setting. In this way, the matrix remains near the manifold of orthogonal matrices after each update. In practice, we have observed that the eigenvalues of $\boldsymbol{W}_{align}$ all have a modulus close to 1, as assumption. Since most of the backdoor image-text pairs contain only a few trigger classes and the provided embedding space closely transforms with the structure of the original embedding space, our watermark injection process satisfies the requirement of retaining downstream task performance without inference in the image-text embedding transformation.

\noindent\textbf{Backdoor Embedding Transformation  Injection}. We compute the provided embedding space $\mathbf{E}_p$ by transforming the original embedding space $\mathbf{E}_o$ to a target embedding space characterized by the pre-defined similarity of triggers and user-specific transformation rules, accomplished through the transformation $\boldsymbol{W}_{align}$ as follows,
\begin{equation}
\begin{aligned}
 \mathbf{E}_p = \boldsymbol{W}_{align}\mathbf{E}_o = \{\boldsymbol{W}_{align}\mathbf{E}_{o_i}, \boldsymbol{W}_{align}\mathbf{E}_{o_s}\}
\end{aligned}
\label{f:injection}
\end{equation}

Since the steals utilize $ \mathbf{E}_p$ for model extraction attack, the learnt transformation matrix $\boldsymbol{W}_{align}$ will serve as a transformation backdoor, which serves as the provider's specific commitment to measure the embedding distribution similarity between the stealer's and the provider's original EaaS for copyright verification, which will be introduced in the copyright verification stage. 

\subsubsection{Copyright Verification.}
\label{sec::verification}
Once a stealer offers a similar service to the public, the EaaS provider can leverage the pre-defined backdoor to verify potential copyright infringement. We propose a collaborative copyright verification strategy by combining backdoor triggers and embedding distribution verification to enhance the watermark robustness.

\noindent\textbf{Backdoor Trigger Verification}. First, we construct a benign image-text set $D'_{c}=\left\{\left[(i_1, s_1), \ldots, (i_m, s_m)\right] \mid(i_k, s_k) \notin D_T\right\}$. Then, we use the image-text pairs from $D'_c$ and the trigger set $D_T$ to query both the stealer and provider models to obtain the stealer's embedding space $\mathbf{E}_{s}=\{(\widetilde{\mathbf{v}}_{k}, \widetilde{\mathbf{t}}_{k})=S_{a}\left(i_{k}, s_{k}\right) \mid (i_{k}, s_{k}) \in D'_{c}+D_T\}$ and original embedding space $\mathbf{E}_{o}=\{(\mathbf{v}_{k}, \mathbf{t}_{k})=S_{v}\left(i_{k}, s_{k}\right) \mid (i_{k}, s_{k}) \in D'_{c}+D_T\}$, respectively. If the similarity of image-text pairs from the trigger set is higher in the stealer's embedding space compared to the original embedding space, we believe that the stealer violates the copyright. 
 To test whether the above conclusion is valid, we first calculate cosine similarity and the square of $\ell_{2}$ distance between trigger image-text pair embeddings in $\mathbf{E}_s$ and $\mathbf{E}_o$,
\begin{equation}
\begin{aligned}
\cos _{i}=\frac{\mathbf{v}_{i} \cdot \mathbf{t}_{i}}{\left\|\mathbf{v}_{i}\right\|\left\|\mathbf{t}_{i}\right\|},& l_{2 i}=\left\|\frac{\mathbf{v}_{i}}{\left\|\mathbf{v}_{i}\right\|}-\frac{\mathbf{t}_{i}}{\left\|\mathbf{t}_{i}\right\|}\right\|^{2}, \\
C_{s}=\left\{\cos _{i} \mid \mathbf{e}_i \in \mathbf{E}_{s}, i \in D_T \right\},& C_{o}=\left\{\cos _{i} \mid \mathbf{e}_i \in \mathbf{E}_{o}, i \in D_T \right\}, \\
L_{s}=\left\{l_{2 i} \mid \mathbf{e}_i \in \mathbf{E}_{s}, i \in D_{T}\right\},& L_{o}=\left\{l_{2 i} \mid \mathbf{e}_i \in \mathbf{E}_{o}, i \in D_{T}\right\} .
\end{aligned}
\label{f:sim}
\end{equation}
\noindent where $\mathbf{e}_i$ donates the embeddings of image-text pair $(\mathbf{v}_i, \mathbf{t}_i)$. Then, we evaluate the detection performance with two metrics of averaged cos similarity and the averaged square of $\ell_{2}$ distance as follows,
\begin{equation}
\begin{aligned}
\Delta_{c o s} & =\frac{1}{\left|C_{s}\right|} \sum_{i \in C_{s}} i-\frac{1}{\left|C_{o}\right|} \sum_{j \in C_{o}} j, \\
\Delta_{l 2} & =\frac{1}{\left|L_{s}\right|} \sum_{i \in L_{s}} i-\frac{1}{\left|L_{o}\right|} \sum_{j \in L_{o}} j .
\end{aligned}
\label{f:sim_delta}
\end{equation}

Since all embeddings are normalized, the ranges of $\Delta_{cos}$ and $\Delta_{l2}$ are [-2,2] and [-4,4], respectively. We report the p-value of the Kolmogorov-Smirnov (KS) test \cite{berger2014kolmogorov} as the third metric, which is used to compare the distribution of $C_s$ and $C_o$. A lower p-value means stronger evidence in favor of the alternative hypothesis.

\noindent\textbf{Embedding Distribution Verification}. We first recover $\mathbf{E}_s$ to pseudo original embedding space $\mathbf{E}'_{o}$ based on $\boldsymbol{W}_{align}$ as follows,
\begin{equation}
\begin{aligned}
 \mathbf{E}'_{o} = \boldsymbol{W}^{-1}_{{align}}\mathbf{E}_s
\end{aligned}
\label{f:recover}
\end{equation}

Then, we evaluate the detection performance by computing the average cosine similarity $C_{avg}$ between $\mathbf{E}'_{o}$ and $\mathbf{E}_o$, as follows,
\begin{equation}
\resizebox{0.92\hsize}{!}{$
\begin{aligned}
\cos _{i}=\frac{\mathbf{e}_{i} \cdot \widetilde{\mathbf{e}}_{i}}{\left\|\mathbf{e}_{i}\right\|\left\|\widetilde{\mathbf{e}}_{i}\right\|},
C_{avg}=\operatorname{Avg}(\left\{\cos _{i} \mid \mathbf{e}_i \in \mathbf{E}_{o}, \widetilde{\mathbf{e}}_i \in \mathbf{E}'_{o},  i \in D'_c \right\}) \\
\end{aligned}$}
\label{f:sim_id}
\end{equation}
\noindent where $\operatorname{Avg}(\cdot)$ represents the average function. We utilize a threshold of $0.4$ in the average $\Delta_{cos}$ and $C_{avg}$ to identify instances of copyright infringement. Furthermore, for multiple users, we calculate the $k$-th average cosine similarity $C_{avg_k}$ and designate the $\boldsymbol{W}_{{align}}$ associated with the highest $C_{avg_k}$ value as the user for user identification.

\section{Experiments}

\subsection{Dataset and Experimental Settings}
Since our focus is to study the impact of our watermark method on the original CLIP-based VLPs model, for this purpose, we conducted zero-shot experiments on 6 datasets: MS-COCO \cite{10.1007/978-3-319-10602-1_48}, Flicker30k \cite{plummer2015flickr30k}, CIFAR-10 \cite{krizhevsky2009learning}, CIFAR-100 \cite{krizhevsky2009learning}, ImageNet-1k \cite{5206848} and VOC2007 \cite{everingham2008pascal}. Specifically, MS-COCO and Flicker30k are widely used datasets for image-text retrieval. CIFAR-10, CIFAR-100, ImageNet-1K, and VOC2007 are large datasets for object classification. Details on each dataset for the zero-shot evaluation and the metrics are provided in Table \ref{tab:datasets}. Additionally, we use the Visual Genome \cite{10.1007/s11263-016-0981-7} dataset with 108,000 samples to count class frequencies. To validate the effectiveness of our watermark detection algorithms, we report three metrics, \textit{i.e.,} the difference of cosine similarity, the difference of squared $\ell_{2}$ distance and p-value (defined in Section \ref{sec::verification}).

\begin{table}[t]
\centering
\caption{Statistics of datasets for zero-shot evaluation.}
\scalebox{1.35}
{\scriptsize
\setlength{\tabcolsep}{2.8mm}
\begin{tabular}{lccc}
\toprule
\multicolumn{1}{l}{Dataset} & \multicolumn{1}{l}{Classes} & \multicolumn{1}{l}{Test size} & \multicolumn{1}{l}{Evaluation metric} \\ \midrule
MS-COCO                      & -                           & 5,000                         & Recall                                \\
Flicker30k                  & -                           & 1,000                         & Recall                                \\ \midrule
CIFAR-10                    & 10                          & 10,000                        & Accuracy                              \\
CIFAR-100                   & 100                         & 10,000                        & Accuracy                              \\
ImageNet                    & 1000                        & 50,000                        & Accuracy                              \\
VOC 2007             & 20                          & 4,952                         & 11-point mAP                          \\ \bottomrule
\end{tabular}}

\label{tab:datasets}
\end{table}

We utilized Adam \cite{kingma2014adam} for training linear transformation $\boldsymbol{W}_{align}$ with a learning rate of $1\times10^{-5}$ and $50,000$ epochs on $1$ Tesla V100 32G GPU in less than $30$ seconds. We initialize the original CLIP model using the CLIP$_{large}$ checkpoints \cite{pmlr-v139-radford21a}, and we conducted inference using the original CLIP model to obtain image-text embeddings as the original EaaS embeddings. The maximum trigger class in a single image is $3$, the size of the trigger class set is $26$, and the size of the trigger image-text set is $1024$. We conducted each experiment independently $5$ times and reported the average results. Moreover, we introduce another threshold $\tau$ to determine instances of copyright infringement. In accordance with established statistical practices, a standard p-value of $5\times10^{-3}$ is considered suitable for rejecting the null hypothesis with statistical significance \cite{benjamin2018redefine}, which can be applied to identify instances of copyright infringement. 

\begin{table*}[t]
\centering
\caption{Performance of different methods on the MS-COCO, Flicker30k, CIFAR-10, CIFAR-100, ImageNet-1K and VOC2007. $\uparrow$ means higher metrics are better. $\downarrow$ means lower metrics are better. R@5 indicates the results of Recall@5 Text / Image Retrieval.}
\scalebox{1.30}
{\scriptsize
\setlength{\tabcolsep}{2.0mm}
\begin{tabular}{lllcccccc}
\toprule
\multicolumn{1}{c}{\multirow{2}{*}{Method}}     & \multicolumn{1}{c}{\multirow{2}{*}{Dataset}} & \multicolumn{1}{c}{\multirow{2}{*}{Metric}} & \multirow{2}{*}{Results (\%)} & \multicolumn{1}{c}{\multirow{2}{*}{$\Delta$}} & \multicolumn{4}{c}{Detection Performance}                            \\ \cmidrule(lr){6-9} 
\multicolumn{1}{c}{}                            & \multicolumn{1}{c}{}                         & \multicolumn{1}{c}{}                        &                               & \multicolumn{1}{l}{}                       & p-value$\downarrow$            & $\Delta_{cos}(\%)\uparrow$                    & $\Delta_{l2}(\%)\downarrow$       & $C_{avg}(\%)\uparrow$               \\ \cmidrule(lr){1-9} 
\multicolumn{1}{c|}{\multirow{6}{*}{Original}}  & MS-COCO                                      & R@5                                         & 91.1 / 89.4                   & \multicolumn{1}{c|}{0.0 / 0.0}             & \multirow{6}{*}{1.00} & \multirow{6}{*}{0.00}   & \multirow{6}{*}{0.00} & \multirow{6}{*}{-}   \\
\multicolumn{1}{l|}{}                           & Flicker30k                                   & R@5                                         & 98.7 / 92.9                   & \multicolumn{1}{c|}{0.0 / 0.0}             &                    &                        &    &                    \\
\multicolumn{1}{l|}{}                           & CIFAR-10                                     & ACC                                       & 96.6                          & \multicolumn{1}{c|}{0.0}                   &                    &                        &       &                 \\
\multicolumn{1}{l|}{}                           & CIFAR-100                                    & ACC                                       & 83.4                          & \multicolumn{1}{c|}{0.0}                   &                    &                        &        &                \\
\multicolumn{1}{l|}{}                           & ImageNet-1K                                  & ACC                                       & 75.4                          & \multicolumn{1}{c|}{0.0}                   &                    &                        &         &               \\
\multicolumn{1}{l|}{}                           & VOC2007                                      & mAP                                         & 84.1                          & \multicolumn{1}{c|}{0.0}                   &                    &                        &        &                \\ \cmidrule(lr){1-9} 
\multicolumn{1}{c|}{\multirow{6}{*}{EmbMarker}} & MS-COCO                                      & R@5                                         & 47.9 / 65.3                            & \multicolumn{1}{c|}{-43.2 / -24.1}                 & \multirow{6}{*}{$<1\times10^{-225}$} & \multirow{6}{*}{51.38}     & \multirow{6}{*}{-53.42} &  \multirow{6}{*}{-}   \\
\multicolumn{1}{l|}{}                           & Flicker30k                                   & R@5                                         & 85.8 / 66.2                            & \multicolumn{1}{c|}{-12.9 / -26.7}                 &                    &                        &                        \\
\multicolumn{1}{l|}{}                           & CIFAR-10                                     & ACC                                       & 77.9                             & \multicolumn{1}{c|}{-18.7}                     &                    &                        &                        \\
\multicolumn{1}{l|}{}                           & CIFAR-100                                    & ACC                                       & 66.8                             & \multicolumn{1}{c|}{-16.6}                     &                    &                        &                        \\
\multicolumn{1}{l|}{}                           & ImageNet-1K                                  & ACC                                       & 62.6                             & \multicolumn{1}{c|}{-12.9}                     &                    &                        &                        \\
\multicolumn{1}{l|}{}                           & VOC2007                                      & mAP                                         & 41.4                             & \multicolumn{1}{c|}{-42.7}                     &                    &                        &                        \\ \cmidrule(lr){1-9}
\multicolumn{1}{c|}{\multirow{6}{*}{Random}} & MS-COCO                                      & R@5                                         & 87.1 / 84.7                              & \multicolumn{1}{c|}{-4.2 / -3.7}                 & \multirow{6}{*}{$< 1\times10^{-81~}$} & \multirow{6}{*}{16.32}     & \multirow{6}{*}{-30.58}    & \multirow{6}{*}{100.00}   \\
\multicolumn{1}{l|}{}                           & Flicker30k                                   & R@5                                         &  98.0 / 87.6                       & \multicolumn{1}{c|}{-0.7 / -5.3}                 &                    &                        &          &              \\
\multicolumn{1}{l|}{}                           & CIFAR-10                                     & ACC                                       & 95.8                             & \multicolumn{1}{c|}{-0.8}                     &                    &                        &                  &      \\
\multicolumn{1}{l|}{}                           & CIFAR-100                                    & ACC                                       & 82.8                             & \multicolumn{1}{c|}{-0.4}                     &                    &                        &                   &     \\
\multicolumn{1}{l|}{}                           & ImageNet-1K                                  & ACC                                       & 75.3                             & \multicolumn{1}{c|}{\textbf{-0.1}}                     &                    &                        &           &             \\
\multicolumn{1}{l|}{}                           & VOC2007                                      & mAP                                         & 80.5                             & \multicolumn{1}{c|}{-3.6}                     &                    &                        &                   &     \\ \cmidrule(lr){1-9}
\multicolumn{1}{c|}{\multirow{6}{*}{VLPMarker}}      & MS-COCO                                      & R@5                                         & 91.1 / 89.2                   & \multicolumn{1}{c|}{\textbf{0.0 / -0.2}}           & \multirow{6}{*}{$< 1\times10^{-186}$} & \multirow{6}{*}{39.06} & \multirow{6}{*}{-1.41} & \multirow{6}{*}{100.00}  \\

\multicolumn{1}{l|}{}                           & Flicker30k                                   & R@5                                         & 98.7 / 92.8                   & \multicolumn{1}{c|}{\textbf{0.0 / -0.1}}             &                    &                        &                &        \\
\multicolumn{1}{l|}{}                           & CIFAR-10                                     & ACC                                       & 96.6                          & \multicolumn{1}{c|}{\textbf{0.0}}                   &                    &                        &                    &    \\
\multicolumn{1}{l|}{}                           & CIFAR-100                                    & ACC                                       & 83.3                          & \multicolumn{1}{c|}{\textbf{-0.1}}                  &                    &                        &                     &   \\
\multicolumn{1}{l|}{}                           & ImageNet-1K                                  & ACC                                       & 75.4                          & \multicolumn{1}{c|}{\textbf{-0.1}}                   &                    &                        &                     &   \\
\multicolumn{1}{l|}{}                           & VOC2007                                      & mAP                                         & 84.1                          & \multicolumn{1}{c|}{\textbf{0.0}}                  &                    &                        &                      &  \\ \bottomrule
\end{tabular}}
\label{tab:results}
\end{table*}

\subsection{Performance Comparison}

\begin{table}[t]
\caption{The impact of the model type.}
\scalebox{1.3}
{\scriptsize
\setlength{\tabcolsep}{0.7mm}
\begin{tabular}{cccccc}
\toprule
\multirow{2}{*}{CLIP}      & \multirow{2}{*}{Parameters} & \multicolumn{4}{c}{Detection Performance} \\ \cmidrule(lr){3-6} 
                           &                             & p-value$\downarrow$       & $\Delta_{cos}(\%)\uparrow$      & $\Delta_{l2}(\%)\downarrow$  & $C_{avg}(\%)\uparrow$    \\ \midrule
\multicolumn{1}{c|}{Base}  & \multicolumn{1}{c|}{82M}    & $<1\times10^{-75~}$        & 31.80        & -0.54      & 100.00 \\
\multicolumn{1}{c|}{Large} & \multicolumn{1}{c|}{400M}   & $<1\times10^{-186}$      & 39.06        & -1.41       & 100.00  \\
\multicolumn{1}{c|}{Huge}  & \multicolumn{1}{c|}{1000M}  & $<1\times10^{-129}$       & 32.13        & -0.69      & 100.00  \\ \bottomrule
\end{tabular}}
\label{tab:model}
\end{table}

\begin{table}[t]
\centering
\caption{The impact of trigger number in six datasets.}
\scalebox{1.3}
{\scriptsize
\setlength{\tabcolsep}{0.4mm}
\begin{tabular}{cccccc}
\toprule
\multirow{2}{*}{\begin{tabular}[c]{@{}c@{}}Trigger\\ Num\end{tabular}} & \multirow{2}{*}{\begin{tabular}[c]{@{}c@{}}Avg \\ Results  $\Delta$(\%)\end{tabular}} & \multicolumn{4}{c}{Detection Performance} \\ \cmidrule(lr){3-6} 
                                                                       &                                   & p-value$\downarrow$       & $\Delta_{cos}(\%)\uparrow$      & $\Delta_{l2}(\%)\downarrow$  & $C_{avg}(\%)\uparrow$  \\ \midrule
\multicolumn{1}{c|}{2}                                                 & \multicolumn{1}{c|}{-0.025}        & $<1\times10^{-1225}$     & 84.78        & -0.63     & 97.82 \\
\multicolumn{1}{c|}{8}                                                 & \multicolumn{1}{c|}{-0.001}        & $<1\times10^{-1225}$      & 89.60        & -0.71    & 98.49  \\
\multicolumn{1}{c|}{32}                                                & \multicolumn{1}{c|}{~0.000}        & $<1\times10^{-1225}$     & 91.72        & -1.39     & 100.00 \\
\multicolumn{1}{c|}{64}                                                & \multicolumn{1}{c|}{~0.000}        & $<1\times10^{-315~}$     &  90.47       & -1.45     & 100.00\\
\multicolumn{1}{c|}{128}                                               & \multicolumn{1}{c|}{-0.025}        & $<1\times10^{-282~}$       & 82.94        & -1.44   & 100.00   \\
\multicolumn{1}{c|}{512}                                               & \multicolumn{1}{c|}{-0.038}        & $<1\times10^{-215~}$       & 55.14        & -1.41   & 100.00   \\
\multicolumn{1}{c|}{1024}                                              & \multicolumn{1}{c|}{-0.050}        & $<1\times10^{-186~}$       & 39.06        & -1.41   & 100.00   \\
\multicolumn{1}{c|}{1536}                                              & \multicolumn{1}{c|}{-0.050}        & $<1\times10^{-147~}$       & 28.53        & -0.75   & 100.00   \\ \bottomrule
\end{tabular}}
\label{tab:number}
\end{table}

We compare the performance of VLPMarker with the following baselines: 1) \textbf{Original}, in which does not backdoor the provided embeddings. 2) \textbf{EmbMarker} \cite{peng2023copying}, a method to backdoor in word embeddings, which selects a target embedding as the watermark and embeds it into the embeddings of texts containing moderate-frequency trigger words as the backdoor. We report metrics calculated by the methodology in the original paper. 3) \textbf{Random}, which randomly updates the linear transformation $\boldsymbol{W}_{align}$ during training without employing our orthogonal constraint method.

Table \ref{tab:results} presents the performance results for all methods, and we have made several observations. \textbf{(1) VLPMarkers' performance across various types of downstream tasks remains consistent with the Original baseline.} This is attributed to our VLPMarker operating on both vision and language embeddings and minimizing the impact on the similarity of visual and language embeddings. \textbf{(2) VLPMarker's detection performance is better than random baselines regarding p-value and the difference in cosine similarity.} This is because we maintain $\boldsymbol{W}_{align}$ close to an orthogonal matrix, ensuring isometric transformations within Euclidean space. This minimizes the impact on the original embedding space when injecting backdoors, as evident from our VLPMarker's minimal change in $\ell_{2}$ distance. In contrast, EmbMarker forces many word embeddings to be close to target embedding, significantly altering the Euclidean space of the original embedding. This alteration leads to a pronounced decline in the task performance of CLIP-based VLP models. \textbf{(3) VLPMarker performs better in retrieval tasks than the random baseline.} This advantage is attributable to our VLPMarker's preservation of fine-grained contextual information within the original embedding space through orthogonalization, which is beneficial in retrieval tasks.
\begin{figure*}[t]
    \centering
    \includegraphics[width=0.84\linewidth]{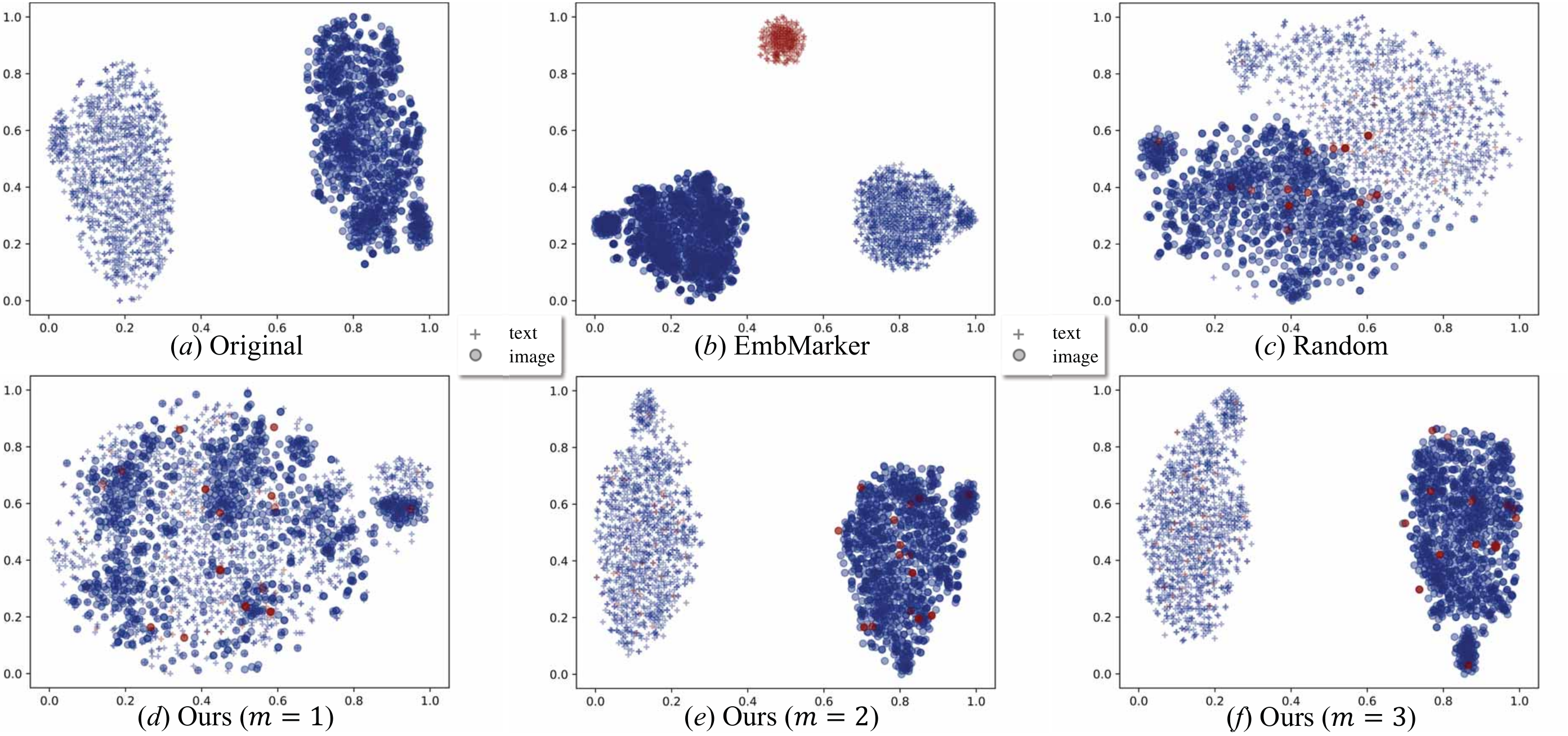}
    \caption{T-SNE visualization of the distribution of provided embedding space of our VLPMarker. ‘\textcolor[RGB]{1,28,127}{Circle}’ means image embeddings, ‘\textcolor[RGB]{107,132,180}{cross}’ indicates text embeddings ‘\textcolor[RGB]{144,48,35}{red}’ means trigger samples.}
    \label{fig:t-SNE}
\end{figure*}
\subsection{Impact of Original Model Type}

To assess the impact of the original model types on VLPMarker's performance, we conducted experiments using different versions of CLIP, including the base, large, and huge models, as the provider model. The same trigger set containing $1024$ image-text pairs was used to train $\boldsymbol{W}_{align}$. As shown in Table \ref{tab:model}, we observe that VLPMarker effectively verifies copyright for all types of provider models. This demonstrates the model-agnostic nature of our VLPMarker, as it operates on the model's embedding space rather than the original model itself. Consequently, VLPMarker can be easily applied to various CLIP-based VLP models.

\subsection{Impact of Trigger Number}

In this section, we assess the impact of the number of triggers on 6 datasets in Table \ref{tab:datasets}. We present the differences in average results across these datasets for different trigger numbers with detection performance in Table \ref{tab:number}, where we can have several observations. \textbf{(1) A small-sized trigger set ($<$ 32) simultaneously reduces model and detection performance.} This is because a small-sized trigger set leads to overfitting of $\boldsymbol{W}_{align}$ to the image-text pairs, which impacts the original embedding space. \textbf{(2) A moderate-sized trigger set (between 32 and 64) has little impact on model performance while exhibiting high detection performance}, indicating limited influence on the original embedding space but affects backdoor inheritance (discussed in Section \ref{sec::Extraction}). \textbf{(3) Results with an appropriate trigger set ($>$ 512) have minimal impact on model performance while still ensuring adequate detection performance.} This is attributed to our orthogonal updating strategy to avoid overfitting many trigger image-text pairs.

\subsection{Embedding Distribution Visualization}
In this section, we focus on assessing the confidentiality of backdoored embeddings to potential stealers. To achieve this, we employ t-SNE \cite{van2008visualizing} for visualizing the embedding space produced by EmbMarker and various settings of our VLPMarker. We present the results in Figure \ref{fig:t-SNE}, where We have several observations. First, the provided embedding spaces produced by our VLPMarker, utilizing trigger image-text pairs featuring combinations of $2\sim3$ high-frequency classes, exhibit consistency with the original embedding space. This transformation can be seen as a `rotation' of the original embedding space, highlighting our VLPMarker's ability to preserve the Euclidean space of the original embedding. Second, backdoored embeddings incorporating triggers exhibit distributions similar to those of benign embeddings, demonstrating the watermark confidentiality of our VLPMarker. Finally, our method employs trigger image-text pairs with a single high-frequency class that embeds watermarks into many provided embeddings, affecting the original embedding space, similar to EmbMarker and Random baselines.

\subsection{Defending Against Attacks}
\label{sec::attack}

We consider model extraction attacks, as defined in Section \ref{sec::extraction}, and similarity-invariant attacks, where the stealer applies similarity-invariant transformations on the copied embeddings \cite{peng2023copying}. 

\subsubsection{Model Extraction Attacks}
\label{sec::Extraction}
In our experiments, the stealer employs OPENCLIP \cite{ilharco_gabriel_2021_5143773}, which is a public version replicating CLIP, as the backbone model to extract the OpenAI CLIP \cite{pmlr-v139-radford21a} as the victim model. We assume that the attacker utilizes the mean squared error (MSE) loss to extract the victim model and uses the Conceptual Caption dataset, which comprises $3$M image-text pairs, as $D_c$, more details are in Appendix \ref{sec:Attacker_imple}. The loss function is defined as follows,
\begin{equation}
\begin{aligned}
\mathbf{\Theta}_{a}^{*}=\arg \min _{\boldsymbol{\Theta}_{a}} \mathbb{E}_{x \in D_{c}}\left\|g\left(x ; \boldsymbol{\Theta}_{a}\right)-\mathbf{e}_{p}^{x}\right\|_{2}^{2}
\end{aligned}
\label{f:steal}
\end{equation}
\noindent where $\mathbf{e}_{p}^{x}$ is the provided embedding of sample $x$ and $g$ is the function of the extracted model. The results of our experiments are presented in Table \ref{tab:extract}, where we have several observations. \textbf{(1) In models `1-3', VLPMarker exhibits superior detection performance compared to EmbMarker.} This is attributed to utilizing multiple high-frequency classes in the trigger image-text pairs. Each predefined combination of trigger classes within a query image-text pair brings the copied image embeddings closer to the predefined copied text embeddings. Consequently, an image featuring multiple predefined combinations of trigger classes results in a copied image embedding closely resembling the predefined copied text embeddings. In contrast, the significant decline in detection performance in EmbMarker is attributed to the scarcity of moderate-frequency words in the image-text pairs. \textbf{(2) In models `4-5', the choice of the trigger number significantly influences detection performance.} A smaller trigger number yields poor detection performance due to the implication of fewer predefined class combinations, thereby reducing the probability of the backdoor being inherited by the stealer model. Conversely, an excessive trigger number increases the inheritance of the backdoor into the recipient model but leads to underfitting of $\boldsymbol{W}_{align}$ as a result of our orthogonal updating strategy, ultimately causing a decrease in detection performance. \textbf{(3) In models `6-7', a trigger set containing moderate-frequency classes or four classes decreases detection performance.} This is attributed to a few image-text pairs containing lower-frequency or complex combinations of trigger classes. Consequently, the inheritance of the backdoor is affected due to the limited predefined class combinations present in the duplicate dataset. \textbf{(4) Our VLPMarker consistently achieves $C_{avg}(\%)$ exceeding 90\% in various settings, enabling robust detection of copyright infringement even with few trigger inheritance. } This is attributed to VLPMarker injecting the embedding space backdoor for all copied embeddings during the transformation.

\begin{table}[t]
\caption{The performance of all methods to defend against model extraction attacks.}
\scalebox{1.25}
{\scriptsize
\setlength{\tabcolsep}{1.0mm}
\begin{tabular}{llcccc}
\toprule
\multirow{1}{*}{} & \multirow{2}{*}{Method}        & \multicolumn{4}{c}{Detection Performance}                \\ \cmidrule(lr){3-6}
                        &       & p-value$\downarrow$       & $\Delta_{cos}(\%)\uparrow$      & $\Delta_{l2}(\%)\downarrow$            & $C_{avg}(\%)\uparrow$                               \\ \midrule
\multicolumn{1}{l}{1.} & \multicolumn{1}{l|}{EmbMarker} & $>1\times10^{-4}$ & 2.19  & \multicolumn{1}{c}{-4.37}  & -                              \\
\multicolumn{1}{l}{2.} & \multicolumn{1}{l|}{Random}    & $<1\times10^{-6}$    & 6.54  & \multicolumn{1}{c}{\textbf{-13.02}} & 99.24                             \\
\multicolumn{1}{l}{3.} & \multicolumn{1}{l|}{VLPMarker} & $<1\times10^{-8}$    & \textbf{11.60} & \multicolumn{1}{c}{-0.72}  & \textbf{99.81}                             \\ \cmidrule(lr){1-6}
\multicolumn{1}{l}{4.} & \multicolumn{1}{l|}{64 Triggers}   & $<1\times10^{-6}$   & 5.05  & \multicolumn{1}{c}{-0.21}  & 96.17                             \\
\multicolumn{1}{l}{5.} & \multicolumn{1}{l|}{1536 Triggers} & $<1\times10^{-8}$    & 9.61  & \multicolumn{1}{c}{-0.13}  & 99.64                             \\ 
\multicolumn{1}{l}{6.} & \multicolumn{1}{l|}{Moderate}    & $<1\times10^{-5}$ & 3.82  & \multicolumn{1}{c}{-0.27}  & 91.87                             \\
\multicolumn{1}{l}{7.} & \multicolumn{1}{l|}{4 Classes}    & $<1\times10^{-7}$ & 8.74  & \multicolumn{1}{c}{-0.52}  & 98.70                            \\
\bottomrule
\end{tabular}}
\label{tab:extract}
\end{table}

\begin{table}[t]
\caption{The performance of all methods to defend against dimension-shift attacks.}
\scalebox{1.25}
{\scriptsize
\setlength{\tabcolsep}{2.8mm}
\begin{tabular}{lccc}
\toprule
\multirow{2}{*}{Method}        & \multicolumn{3}{c}{Detection Performance} \\ \cmidrule(lr){2-4} 
                               & p-value$\downarrow$       & $\Delta_{cos}(\%)\uparrow$      & $\Delta_{l2}(\%)\downarrow$       \\ \midrule
\multicolumn{1}{l|}{EmbMarker} & $>1\times10^{-3}$   & 1.82    & -3.28    \\
\multicolumn{1}{l|}{Random}    & $<1\times10^{-6}$      & 6.54    & \textbf{-13.02}   \\
\multicolumn{1}{l|}{VLPMarker}      & $<1\times10^{-8}$   & \textbf{11.60}   & -0.72  \\ \bottomrule
\end{tabular}}
\label{tab:shift}
\end{table}

\subsubsection{Similarity-invariant Attacks} Following \cite{peng2023copying}, we denote similarity invariance as below.
\label{sec::sim}

\begin{define}
($l$ Similarity Invariance).
For a transformation $\mathbf{A}$, given every vector pair $(\textbf{i}, \textbf{j})$, $\mathbf{A}$ is $l$-similarity-invariant only if $l(\mathbf{A}(\textbf{i}), \mathbf{A}(\textbf{j})) = l(\textbf{i}, \textbf{j})$, where $l$ is a similarity metric.
\end{define}

We assume the stealer applies similarity-invariant attacks, such as dimension-shift attacks, after gaining access to the stolen model through model extraction. The experimental results of all methods under dimension-shift attacks are presented in Table \ref{tab:shift}. Notably, the detection performance of VLPMarker remains consistent under dimension-shift attacks, providing high confidence in concluding that the stealer has violated the copyright of the EaaS provider, even in cases where our embedding distribution verification becomes ineffective. This is because VLPMarker injects the backdoor of predefined image embeddings closer to their corresponding predefined text embeddings, which the stealer's model effectively inherits. Consequently, for any similarity-invariant attacks, the predefined image embeddings remain close to the corresponding predefined text embeddings. In contrast, EmbMarker uses text embeddings from the provider's model as a target embedding to combat such attacks, decreasing detection performance. We have theoretically demonstrated that their detection performance should remain the same for other similarity-invariant attacks in Appendix \ref{sec:sim_prove}.

\subsection{User Identification}

\begin{table}[t]
\centering
\caption{Identification performance of different methods.}
\scalebox{1.3}
{\scriptsize
\setlength{\tabcolsep}{1.8mm}
\begin{tabular}{lcllllll}
\toprule
\multirow{2}{*}{Method}        & \multirow{2}{*}{User Amount} & \multicolumn{2}{c}{Identification Performance} \\ \cmidrule(lr){3-4} 
                               &                             & Avg Cos (\%)           & Accuracy (\%)           \\ \midrule
\multicolumn{1}{l|}{EmbMarker} & 1024                        & 83.627                 & 92.659             \\
\multicolumn{1}{l|}{Random}    & 1024                        & 100.00                 & 100.00             \\
\multicolumn{1}{l|}{VLPMarker}      & 1024                        & 100.00                 & 100.00             \\ \bottomrule
\end{tabular}}
\label{tab:id}
\end{table}

\begin{figure}[t]
    \centering
    \centering
    \includegraphics[width=1.0\linewidth]{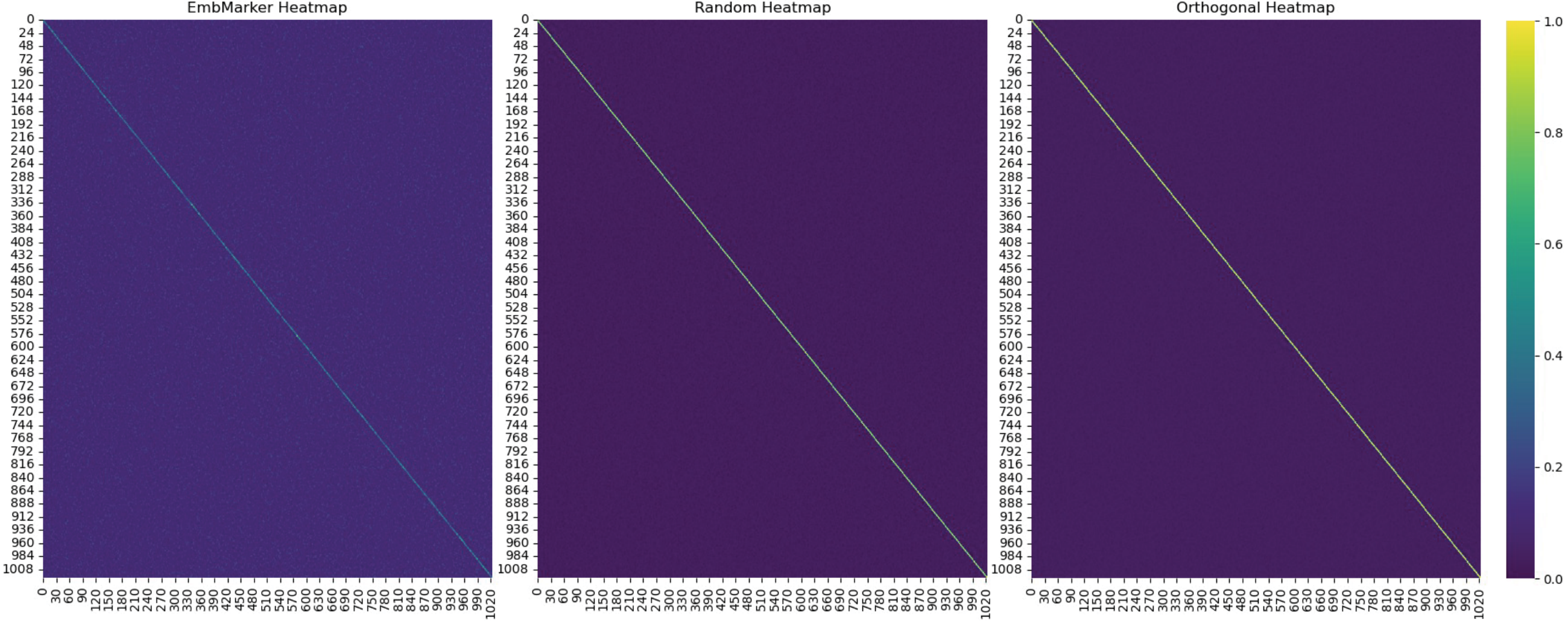}
    \caption{Similarity visualization of different methods.}
    \vspace{-6pt}
    \label{fig:heat}
\end{figure}

Multi-model EaaS deployed in multi-user scenarios, and our VLPMarker can generate different $\boldsymbol{W}_{{align}_k}$ to transform the same original embedding space to unique embedding spaces for different users, facilitating user identification. We consider a scenario with $1024$ users and employ the identification method discussed in Section \ref{sec::verification} and Appendix \ref{sec:id_setting}. Our results are summarized in Table \ref{tab:id}, where we have several observations. First, VLPMarker outperforms EmbMarker in user identification. This is attributed to the uniqueness of our $\boldsymbol{W}_{{align}_k}$, which transforms the same original embedding space to a user-specific provided embedding space, creating robust discriminative features. Second, the performance of the random baseline is consistent with VLPMarker, which indicates that our treating the watermark injection process as an embedding space transform process is a robust and reliable method. Furthermore, in Figure \ref{fig:heat}, we visualize the cosine similarity matrix for different methods. VLPMarker's diagonal elements exhibit a noticeable contrast in values compared to non-diagonal elements, proving the significant discriminative ability of VLPMarker's provided embedding spaces.

\section{Conclusion}

In this paper, we proposed VLPMarker, a novel watermarking method for CLIP-based VLPs on multi-modal EaaS. VLPMarker achieves the injection of backdoor triggers into VLPs through transformation, ensuring efficient watermarking without requiring fine-tuning of the models. Our VLPMarker minimizes the impact on model performance while preserving the original transform structure between visual and language embeddings. Our copyright verification strategy enhances the watermark's resilience against various attacks. Notably, VLPMarker uses OoD triggers, eliminating the need for access to the model training data, making it applicable to real-world scenarios. How to design a more efficient transformation-based watermarking method will be the future work.

\bibliographystyle{ACM-Reference-Format}
\bibliography{sample-base}

\clearpage

\appendix
\section{Experimental Settings}

\subsection{Attacker Model Implementation}
\label{sec:Attacker_imple}
We adopt ViT-L/14 OpenAI CLIP \cite{pmlr-v139-radford21a} pre-trained on 400M image-text paired data as the provider model, ViT-L/14 OPENCLIP\cite{ilharco_gabriel_2021_5143773} as the stealer model. For training the stealer model, we utilize the Conceptual Caption dataset \cite{DBLP:conf/acl/SoricutDSG18}, which comprises 3M images. We employ AdamW \cite{loshchilov2018decoupled} with a learning rate of $1\times10^{-3}$, weight decay of $0.1$, and a linear warmup of $10000$ steps. The batch size is $64$. Our model is trained on $4$ Tesla V100 (32G) GPUs for $50$ epoch.

\subsection{User Identification Settings}
\label{sec:id_setting}
We considered a scenario with $1024$ users, using a trigger set with $1024$ image-text pairs, generating unique $\boldsymbol{W}_{{align}_k}$ for different users, providing user-specific embedding spaces with $1024$ random image-text pairs and identify each user, which first recover $\mathbf{E}_s$ to different $\mathbf{E}'_{o_k}$ based on each $\boldsymbol{W}_{align}$ as follows,
\begin{equation}
\begin{aligned}
 \mathbf{E}'_{o_k} = \boldsymbol{W}^{-1}_{{align}_k}\mathbf{E}_s
\end{aligned}
\label{f:recover}
\end{equation}
where $\boldsymbol{W}_{{align}_k}$ is the user-special transformation rule of $k$-th user. Then, we compute the  $k$-th average cosine similarity $C_{avg_k}$ between for $\mathbf{E}'_{o_k}$ and $\mathbf{E}_o$, as illustrated below:
\begin{equation}
\begin{aligned}
\cos _{i}=\frac{\mathbf{e}_{i} \cdot \widetilde{\mathbf{e}}_{i}}{\left\|\mathbf{e}_{i}\right\|\left\|\widetilde{\mathbf{e}}_{i}\right\|},
C_{avg_k}=\operatorname{Avg}(\left\{\cos _{i} \mid \mathbf{e}_i \in \mathbf{E}_{o}, \widetilde{\mathbf{e}}_i \in \mathbf{E}'_{o},  i \in D'_c \right\}) \\
\end{aligned}
\label{f:sim_id}
\end{equation}
\noindent where $\operatorname{Avg}(\cdot)$ denotes average function. Finally, we identify the index of $\boldsymbol{W}_{{align}_k}$ associated with the highest value of $C_{avg_k}$ as the User.

\section{Theoretical Proof}
\label{sec:sim_prove}
In this section, we provide theoretical proof in Section~\ref{sec::sim}.

\begin{prop}
Denote identity transformation $\mathbf{I}$ as $\mathbf{I}(\textbf{e}) = \textbf{e}$ and dimension-shift transformation $\mathbf{S}$ as
$\mathbf{S}(\textbf{e}) = (e_d, e_1, e_2, \dots, e_{d - 1})$,
where $\textbf{e}$ is a embedding, $e_i$ is the $i$-th dimension of $\textbf{e}$ and $d$ is the dimension of $\textbf{e}$.
Both identity transformation $\mathbf{I}$ and dimension-shift transformation $\mathbf{S}$ are similarity-invariant.
\label{prop:sim-inv}
\end{prop}

\begin{prop}
For a copied model, the detection performance $\Delta_{cos}$, $\Delta_{l2}$ and p-value of our VLPMarker remains consistent under any two similarity-invariant attacks involving transformations $\mathbf{A}_1$ and $\mathbf{A}_2$, respectively.
\label{prop:same-perf}
\end{prop}

\subsection{Proof of Proportion 1}
\label{sec:ds-invariant}

\textit{Proof.}
Given any pair of image-text embedding $(\textbf{v}, \textbf{t})$, according to the definition of identity transformation, we have 

\begin{equation}
\begin{aligned}
        &||\frac{\mathbf{I}(\textbf{v})}{||\mathbf{I}(\textbf{v})||} - \frac{\mathbf{I}(\textbf{t})||^2}{||\mathbf{I}(\textbf{t})||} = ||\frac{\textbf{v}}{||\textbf{v}||} - \frac{\textbf{t}}{||\textbf{t}||}||_2^2, \\
        &cos(\mathbf{I}(\textbf{v}), \mathbf{I}(\textbf{t})) = cos(\textbf{v}, \textbf{t}),
\end{aligned}
\nonumber
\end{equation}
which indicates identity transformation is similarity-invariant.

For dimension-shift transformation $\mathbf{S}$, we have
\begin{equation}
\begin{aligned}
    &||\frac{\mathbf{S}(\textbf{v})}{||\mathbf{S}(\textbf{v})||} -\frac{\mathbf{S}(\textbf{t})}{||\mathbf{S}(\textbf{t})||}||^2 \\
    &= \sum_{k=1}^{d} (\frac{v_k}{||\textbf{v}||} - \frac{t_k}{||\textbf{t}||})^2
    = ||\frac{\textbf{v}}{||\textbf{v}||} - \frac{\textbf{t}}{||\textbf{t}||}||^2,
\end{aligned}      
\nonumber
\end{equation}

\begin{equation}
\begin{aligned}
     cos(\mathbf{S}(\textbf{v}), \mathbf{S}(\textbf{t})) &= \frac{\sum_{k=1}^{d} v_k t_k}{||\textbf{v}|| \,||\textbf{t}||}
    =  cos(\textbf{v}, \textbf{j}),
\end{aligned}      
\nonumber
\end{equation}
where $d$ is the dimension of $\textbf{v}$ and $\textbf{t}$. Therefore, dimension-shift transformation $\mathbf{S}$ is similarity-invariant as well.

\subsection{Proof of Proportion 2}
\label{sec:defend-all-ds}
\textit{Proof.} Given any pair of image-text embedding $(\textbf{v}, \textbf{t})$, the embedding manipulated by transformation $\mathbf{A}_1$ as $(\textbf{v}^1, \textbf{t}^1)$ and the the embedding manipulated by transformation $\mathbf{A}_2$ as $(\textbf{v}^2, \textbf{t}^2)$.
Since both $\mathbf{A}_1$ and $\mathbf{A}_2$ are similarity-invariant, we have 
\begin{equation}
    \begin{aligned}
    &cos_i^1 = cos_i^2 = cos_i = \frac{\textbf{v}_i \cdot \textbf{t}_i}{||\textbf{v}_i||\, ||\textbf{t}_i||}, \\
    &l_{2i}^1 = l_{2i}^2 = l_{2i} = ||\textbf{v}_i/||\textbf{v}_i|| - \textbf{t}_i/||\textbf{t}_i|| \,||^2,
    \end{aligned}
\nonumber
\end{equation}
where the superscript indicates the similarity calculated under which transformation.
Therefore, we can obtain:
\begin{equation}
    \begin{aligned}
    &C_b^1 = C_b^2, C_n^1 = C_n^2, L_b^1 = L_b^2, L_n^1 = L_n^2.
    \end{aligned}
\nonumber
\end{equation}

Since the inputs for the metrics $\Delta_{cos}$, $\Delta_{l2}$ and p-value in our methods are only $C_b$, $C_n$, $L_b$ and $L_n$, we have
\begin{equation}
    \Delta_{cos}^1 = \Delta_{cos}^2, \Delta_{l2}^1 = \Delta_{l2}^2, p_{KS}^1 = p_{KS}^2,
\nonumber
\end{equation}
where $p_{KS}$ is the p-value of the KS test with $C_s$ and $C_o$ as inputs.

\end{document}